\documentstyle[12pt]{article}

\catcode`\@=11
\def\marginnote#1{}
\def\titlepage{\@restonecolfalse\if@twocolumn\@restonecoltrue\onecolumn
     \else \newpage \fi \thispagestyle{empty}\c@page\z@
        \def\thefootnote{\fnsymbol{footnote}} }
\def\endtitlepage{\if@restonecol\twocolumn \else  \fi
        \def\thefootnote{\arabic{footnote}}
        \setcounter{footnote}{0}}  
\catcode`@=12
\relax
\def\bea{\begin{array}}
\def\beq{\begin{equation}}
\def\eea{\end{array}}
\def\eeq{\end{equation}}

\def\Im{\mathop{\rm Im}}
\def\NP#1#2#3{Nucl. Phys. \underline{#1} (19#2) #3}
\def\ov{\overline}

\def\PL#1#2#3{Phys. Lett. \underline{#1} (19#2) #3}
\def\PR#1#2#3{Phys. Rev. \underline{#1} (19#2) #3}

\def\Re{\mathop{\rm Re}}
\def\Tr{\mathop{\rm Tr}}

\relax

\def\crbig{\\\noalign{\vspace {3mm}}}
\def\bigint{{\displaystyle\int}}
\def\Fint{{\displaystyle{\int d^2\theta\,}}}
\def\Dint{{\displaystyle{\int d^2\theta d^2\ov\theta\,}}}
\def\Fbarint{{\displaystyle{\int d^2\ov\theta\,}}}
\relax

\begin{document}
\topmargin-2.4cm
\begin{titlepage}
\begin{flushright}
NEIP--95--011 \\
hep-th/9511036 \\
September 1995
\end{flushright}
\vskip 0.25in

\begin{center}{\Large\bf
Gaugino Condensation with a Linear Multiplet
\footnote{Work supported in part by the European Union (contract
SC1$^*$-CT92-0789).}}
\vskip .2in
{\bf J.-P. Derendinger}
\vskip .1in
Institut de Physique \\
Universit\'e de Neuch\^atel \\
CH--2000 Neuch\^atel, Switzerland
\end{center}

\vskip .2in
\begin{center}
{\bf Abstract}
\end{center}

\begin{quote}
\noindent
An effective lagrangian analysis of gaugino condensation
is performed in a supersymmetric gauge theory
with field-dependent gauge couplings described with a
linear multiplet. An original aspect of this effective lagrangian
is the use of a real
vector superfield to describe composite gauge
invariant degrees of freedom.
The duality equivalence of this approach with the more familiar
formulation using a chiral superfield is demonstrated.
These results strongly
suggest that chiral-linear duality survives nonperturbative effects in
superstrings.
\end{quote}

\vspace{4.5cm}
\begin{center}
{\it
Invited contribution presented at the International Workshop on
Supersymmetry and Unification of Fundamental Interactions, SUSY95,
Ecole Polytechnique, Palaiseau,
France, 15--19 May 1995.}
\end{center}
\end{titlepage}
\setcounter{footnote}{0}
\setcounter{page}{0}
\newpage
%
%

\section{Introduction}

Dynamical gaugino condensation is expected to occur in supersymmetric,
asymptoti\-cal\-ly-free gauge theories. In the super-Yang-Mills system
without matter, expectation values of gaugino condensates
do not break supersymmetry \cite{NW,VY}. But in theories
with field-dependent
gauge couplings, gaugino condensates are a potential
source for dynamical
supersymmetry breaking. This is true in supergravity theories
\cite{FGN} but
also with global supersymmetry. And superstrings offer
maybe the most
interesting application of this phenomenon \cite{DIN,DRSW}.
As demonstrated in ref. \cite{VY},
the effective lagrangian method has proven
simple and efficient in the analysis of the effect of gaugino
condensates on supersymmetry \cite{AKMRV}.

In superstrings, the string gauge coupling is a field-dependent quantity,
which can be described in two equivalent, dual ways. The gravity sector
of superstrings naturally
contains, as partners of the graviton, a real scalar
field, the dilaton, which defines the string coupling,
an antisymmetric tensor $b_{\mu\nu}$, and their
supersymmetric partners. Since the couplings of the antisymmetric tensor
have the gauge symmetry
\beq
\label{bsym}
b_{\mu\nu}\quad\longrightarrow\quad b_{\mu\nu}+ \partial_{[\mu}\Lambda_
{\nu]},
\eeq
a first supersymmetric description of the dilaton multiplet uses
a linear superfield \cite{linear}, $L$. In the global case, $L$ is a
real vector superfield subject to the constraints
\beq
\label{Ldef}
\ov{\cal DD} L = {\cal DD}L =0.
\eeq
Solving them leads to the component expansion
\beq
\label{Lexp1}
L = C + i\theta\chi -i\ov\theta\ov\chi -(\theta\sigma_\mu\ov\theta)
{1\over\sqrt2}\epsilon^{\mu\nu\rho\sigma}\partial_\nu b_{\rho\sigma}
+{1\over2}\theta\theta\ov\theta (\partial_\mu\chi)\sigma^\mu
+{1\over2}\ov{\theta\theta}\theta\sigma^\mu(\partial_\mu\ov\chi)
+{1\over4}\theta\theta\ov{\theta\theta}\Box C.
\eeq
Up to a field redefinition, the real scalar $C$ is the dilaton.
Symmetry (\ref{bsym}) is a consequence of the defining conditions
(\ref{Ldef}): $L$ itself is inert under (\ref{bsym}). Notice the
absence of auxiliary fields: $b_{\mu\nu}$ has three
off-shell components for only one on-shell degree of freedom.

The existence of a second description is a
well-known manifestation of duality.
A field theory with an antisymmetric tensor $b_{\mu\nu}$ and symmetry
(\ref{bsym}) can always be transformed into an equivalent theory
in which a pseudoscalar with axion-like couplings
replaces $b_{\mu\nu}$.
With supersymmetry, this equivalence is the statement that a model with
a linear multiplet can always be transformed into a dual theory where
$L$ is replaced by a chiral superfield ${\cal S}$. This transformation
will be called chiral-linear duality. The dual ${\cal S}$--theory
contains a complex scalar $s$ (instead of $C$ and $b_{\mu\nu}$)
and a Majorana spinor $\psi_s$ replacing $\chi$ as well as a
complex auxiliary field $f_s$. It is invariant under
\beq
\label{Ssym}
{\cal S} \quad\longrightarrow\quad {\cal S} + i\alpha, \qquad\qquad
\alpha:{\rm a\,\,real\,\,constant},
\eeq
a symmetry related to (\ref{bsym}) by chiral-linear duality.

In the context of superstrings,
the existence of chiral-linear duality is well established at the (string)
perturbative level. Its status when nonperturbative effects like gaugino
condensation are included is more ambiguous. One difficulty has to do with
the apparent necessity of preserving symmetry (\ref{Ssym}) to be able to
perform the duality transformation. And (\ref{Ssym}) is apparently
broken by anomalies in a strongly-coupled gauge sector.
The purpose of these notes
is to perform the effective lagrangian analysis of gaugino condensates
in the linear multiplet description of the dilaton sector and to show that
a certain form of chiral-linear duality survives condensation.
In most parts, they follow ref. \cite{BDQQ1}, a work done
with C. P. Burgess,
F. Quevedo and M. Quir\'os. For simplicity and brevity,
the discussion
will be limited here to global supersymmetry. Its extension
to supergravity is
straightforward, using for instance the formalism of
conformal supergravity with a chiral compensating multiplet $S_0$.

\section{The chiral case}

We start with a summary of the effective lagrangian analysis
of gaugino condensates in the standard case of gauge couplings described
by a single chiral multiplet. We consider the case of a semi-simple
gauge group $G=\prod_a G_a$. We also assume that the theory only contains
gauge-singlet chiral matter multiplets, one of them, $S$, giving the
field-dependent gauge coupling. The lagrangian is then
\beq
\label{LS1}
\begin{array}{rcl}
{\cal L} &=& {\cal L}_S + {\cal L}_G \crbig
{\cal L}_G &=&\displaystyle{{1\over2}\sum_a c_a \Fint S\Tr (W_aW_a)
+ {1\over2}\sum_a c_a \Fbarint \ov S\Tr (\ov W_a\ov W_a),}
\end{array}
\eeq
with real constants $c_a$. ${\cal L}_S$ contains the kinetic terms for
$S$ as well as all contributions of other matter multiplets
which will be omitted below.
${\cal L}_S$ does not depend on the gauge multiplets.
$W_a$ is the spinor chiral superfield of gauge curvatures: $W_{a\alpha} =
-{1\over4}\ov{\cal DD}e^{-V_a}{\cal D}_\alpha e^{V_a}$.
The gauge coupling for group factor $G_a$ is then
\beq
\label{gaugeS}
{1\over g_a^2} = 2c_a \Re s,
\eeq
$s$ being the complex scalar component of the superfield $S$.
Gaugino bilinears appear in the lowest component of the expansion of the
chiral superfields $W_aW_a$,
$$
W_aW_a = -\lambda_a\lambda_a + \ldots.
$$
An expectation value of these bilinears will not, in general, break
supersymmetry.

To obtain the effective lagrangian for gaugino condensates, we
follow ref.
\cite{CJT}\footnote{
For a more complete discussion, see \cite{BDQQ2}.
} and consider the generating functional of two-particle irreducible
(2PI) Green's functions for the gauge sector,
\beq
\label{2PI}
{\rm exp}\left\{i\hat W[J_a,S]\right\} =
\int {\cal D A}\, {\rm exp}\,i\left\{\sum_a \int d^4x\Fint \left({1\over2}
c_aS+J_a \right) \Tr(W_aW_a) +{\rm h.c.} \right\},
\eeq
with chiral external currents $J_a$ coupled to the operators
$\Tr(W_aW_a)$ which contain the bilinears we are interested in.
${\cal D A}$ denotes the integration over all gauge multiplets $V_a$.
This expression factorises in the
absence of charged matter. A Legendre transformation from the variables
$J_a$ to the chiral variables $U_a$ leads to the 2PI effective action
\beq
\label{Legendre}
\Gamma[U_a,S] = \hat W[J_a,S] - \sum_a\int d^4x\, \left[\Fint U_aJ_a
+{\rm h.c.} \right],
\eeq
where the chiral superfields $U_a$ are given by
\beq
\label{Uadef}
U_a = {\delta\hat W\over\delta J_a} = \langle\Tr(W_aW_a)\rangle.
\eeq
Their dimension is three. Clearly,
$$
\Gamma[U_a,S] = \sum_a \Gamma_a[U_a,S],
$$
and the contribution of each simple gauge group factor is of the form
\beq
\label{LUa}
\begin{array}{rcl}
\Gamma_a[U_a,S] &=& \bigint d^4x\, {\cal L}_{U_a}, \crbig
{\cal L}_{U_a} &=& \Dint K_{U_a}(U_a,\ov U_a) +
\Fint w_{U_a}(U_a) + \Fbarint \ov w_{U_a}(\ov U_a).
\end{array}
\eeq
Finally, the effective lagrangian for gaugino bilinears is
\beq
\label{LS2}
{\cal L}_{eff.} = {\cal L}_S + \sum_a {\cal L}_{U_a}.
\eeq
It is a functional of the classical superfields $U_a$
similar to the (1PI) effective lagrangian and effective
potential commonly used to determine expectation values of scalar
fields.

An expression for the superpotentials has been obtained in refs.
\cite{VY,T}:
\beq
\label{WU}
w_{U_a}(U_a) = {1\over2}c_aSU_a +{A_a\over12}U_a\log
\left(U_a\over M^3\right),
\qquad A_a = {3C(G_a)\over 8\pi^2}.
\eeq
The first term will play an important role in the
discussion of duality with
gaugino condensates. The form of the second will
only be used when
discussing the physics of the effective theory.
As usual, the explicit
knowledge of the K\"ahler potentials $K_{U_a}$ is
not of first importance for
the analysis of the vacuum structure implied by
the effective potential. It has
been argued in ref. \cite{VY} that $K_{U_a} =
h_a (U_a\ov U_A)^{1/3}$, with
$h_a$ a dimensionless constant, as suggested
by scale covariance of the lagrangian.

It is useful to realize that each chiral
superfield $U_a$ is not entirely
arbitrary. There exists a real vector superfield $V_a$ such that
\beq
\label{Va}
U_a = -{1\over2}\ov{\cal DD}V_a , \qquad
\ov U_a = -{1\over2}{\cal DD}V_a.
\eeq
This result follows from the fact that
\beq
\label{CS1}
\Tr(W_aW_a) = \ov{\cal DD}\Omega_a, \qquad
\Tr(\ov W_a\ov W_a) = {\cal DD}\Omega_a,
\eeq
in terms of the Chern-Simons real vector superfield $\Omega_a$ for group
$G_a$. Eqs. (\ref{2PI}) and (\ref{Uadef}), lead then to result (\ref{Va}).
In components, eq. (\ref{Va}) indicates that the imaginary part of the
highest $\theta\theta$ component $f_a$ of the chiral $U_a$
is a space-time derivative,
\beq
\label{Imfa}
\Im f_a = \partial_\mu v^\mu_a.
\eeq
With eq. (\ref{Uadef}), this is a consequence of
$$
\Im[\Tr(W_aW_a)]_{\theta\theta} = \Tr\left[
{1\over4}\epsilon_{\mu\nu\rho\sigma}F_a^{\mu\nu}F_a^{\rho\sigma}
-\partial^\mu(\lambda_a\sigma_\mu\ov\lambda_a)\right]
=\partial^\mu \Tr\left[ \epsilon_{\mu\nu\rho\sigma}
\omega_a^{\nu\rho\sigma}
-(\lambda_a\sigma_\mu\ov\lambda_a) \right],
$$
where $\omega_a^{\nu\rho\sigma}$ is the bosonic Chern-Simons form.

\section{The linear case}

The Chern-Simons superfield $\Omega_a$ can be used to construct a
supersymmetric gauge theory with field-dependent gauge couplings.
The gauge
variation $\delta\Omega_a$ of $\Omega_a$ is a linear multiplet,
$$
\ov{\cal DD}\,\delta\Omega_a = {\cal DD}\,\delta\Omega_a = 0.
$$
The idea is to introduce a linear multiplet
matter superfield $L$, with,
by definition, ${\cal DD}L=\ov{\cal DD}L=0$,
and to postulate the gauge transformation
$$
\delta L = 2\sum_a c_a\delta\Omega_a
$$
so that the combination
\beq
\label{Lhatdef}
\hat L = L-2\sum_a c_a\Omega_a
\eeq
is a gauge-invariant vector superfield. In addition since
\beq
\left[ \Omega_a \right]_{\theta\theta\ov\theta\ov\theta}=
\Tr\left[ {1\over8}F_a^{\mu\nu}F_{a\,\mu\nu} -{i\over4}\lambda_a
\sigma^\mu D_\mu\ov\lambda_a +{i\over4}D_\mu\lambda_a\sigma^\mu\ov
\lambda_a-{1\over4}D_aD_a \right],
\eeq
the supersymmetric lagrangian
\beq
\label{LL1}
{\cal L}_L = 2\mu^2\Dint \Phi (\hat L/\mu^2) ,
\eeq
is a supersymmetric gauge theory with gauge couplings
\beq
\label{g2}
{1\over g_a^2} = 2c_a\Phi_x, \qquad \Phi_x =
\left[{d\over dx}\Phi(x)\right]_{x=C\mu^{-2}},
\eeq
which depend on the real scalar lowest component $C$
of the linear multiplet
$L$. It is useful to introduce a scale parameter $\mu$
to keep track of the
physical dimension of $\hat L$, which is two.

Our main concern in this section is to establish
the effective lagrangian
for gaugino bilinears in the linear multiplet formalism,
as in the previous section for the chiral case.

In view of the peculiar gauge transformations of $L$ and
$\Omega_a$, it
is not very convenient to work with theory (\ref{LL1}).
An equivalent
theory is obtained by replacing $\hat L$ by an
unconstrained vector superfield
$V$, and then imposing $V= L-2\sum_a c_a\Omega_a$
as an equation of motion.
Consider then the lagrangian
\beq
\label{LL2}
{\cal L}_V = 2\mu^2\Dint \Phi(V/\mu^2) +
\left({1\over4}\Fint {\cal S}\ov{\cal DD}(V+2\sum_ac_a\Omega_a) +{\rm
h.c.}\right),
\eeq
where ${\cal S}$ is a chiral superfield.
The equation of motion for ${\cal S}$, which acts
like a Lagrange multiplier, imposes the constraints
$$
\ov{\cal DD}(V+2\sum_ac_a\Omega_a)={\cal DD}(V+2\sum_ac_a\Omega_a)=0,
$$
which indicate that $V+2\sum_ac_a\Omega_a$ is a linear multiplet $L$,
or $V=\hat L$. Using (\ref{CS1}), theory (\ref{LL2}) becomes
\beq
\label{LL3}
\begin{array}{rcl}
{\cal L}_V &=& \displaystyle{{\cal L}_{V,{\cal S}} +
{1\over2}\sum_a c_a\Fint {\cal S}\Tr(W_aW_a) +
{1\over2}\sum_a c_a\Fbarint\ov {\cal S}\Tr(\ov W_a\ov W_a),}
\crbig
{\cal L}_{V,{\cal S}} &=&
2\mu^2\Dint \Phi(V/\mu^2) + \left(
{1\over4}\Fint {\cal S}\ov{\cal DD}V +{\rm h.c.}\right).
\end{array}
\eeq
${\cal L}_{V,{\cal S}}$ does not depend on the gauge multiplets,
and the gauge terms
are identical with those in the chiral lagrangian (\ref{LS1}). The
effective lagrangian for gaugino bilinears can then be obtained by
repeating the discussion of the previous section, with
${\cal S}$ and ${\cal L}_{V,{\cal S}}$
replacing $S$ and ${\cal L}_S$. One then finds
\beq
\label{LeffL1}
{\cal L}_{eff.}(V,{\cal S},U_a) = 2\mu^2\Dint \Phi(V/\mu^2) +
{1\over4}\Fint {\cal S}\ov{\cal DD}V + {1\over4}\Fbarint \ov
{\cal S}{\cal DD}V
+\displaystyle{\sum_a c_a{\cal L}_{U_a}},
\eeq
with ${\cal L}_{U_a}$ as in eq. (\ref{LUa}).
This effective theory depends,
as in the chiral case, on chiral superfields $U_a$,
one for each simple
factor in the gauge group. It also depends on an unconstrained vector
superfield $V$, and on the chiral ${\cal S}$\footnote{
The use of a vector superfield in the effective lagrangian
for gaugino
condensates with a linear multiplet has also been suggested
in ref. \cite{BGT}.
But this article does not address the issue of duality equivalence.
}. The dependence on ${\cal S}$
in ${\cal L}_{U_a}$ is however very simple: the superpotential $w_{U_a}$
contains a term linear in ${\cal S}$, ${1\over2}c_a{\cal S}U_a$.
The ${\cal S}$-terms in ${\cal L}_{eff.}$ are then:
$$
{1\over4}\Fint {\cal S}\Big( \ov{\cal DD}V +2\sum_a c_aU_a\Big)
+{\rm h.c.},
$$
and ${\cal S}$ is a Lagrange multiplier superfield
imposing the constraint
\beq
\label{const}
\sum_a c_aU_a = -{1\over2}\ov{\cal DD}V.
\eeq
The interpretation of this equation is very simple.
The variables $U_a$ are
related, by the Legendre transformation (\ref{Legendre}),
to $\Tr(W_aW_a)$
[see eq. (\ref{Uadef})]. On the other hand,
it is the role of the
Lagrange multiplier ${\cal S}$
in theory (\ref{LL2}) to relate $V$ to
$\hat L= L-2\sum_a c_a \Omega_a$. And $\ov{\cal DD}\hat L =
-2\sum_a c_a \Tr(W_aW_a)$ immediately leads to the above condition.

In the case of a simple gauge group, with only one $U_a$
(and $c_a=1$) denoted by $U$,
eq. (\ref{const}) can be used to eliminate $U$. The effective lagrangian
for gaugino condensates in this case reduces to
\beq
\label{Leffsimple}
\begin{array}{rcl}
{\cal L}_{eff.}(V) &=& \Dint \left[2\mu^2\Phi(V/\mu^2)
+K_U(U,\ov U)\right] + \Fint w(U) + \Fbarint \ov w(\ov U), \crbig
w(U) &=& {A\over12}U\log (U/M^3), \qquad
K_U=h(U\ov U)^{1/3},\qquad U=-{1\over2}\ov{\cal DD}V.
\end{array}
\eeq
With several group factors, it will be more convenient to analyse
the effective lagrangian using expression (\ref{LL3}), with ${\cal S}$.

\section{Chiral-linear duality}

In the previous sections, we have constructed the
effective lagrangians for
gaugino condensates in the cases of supersymmetric
gauge theories with
field-dependent gauge couplings described with
either a chiral $S$ or a linear $L$
superfield. It is known that a duality transformation
can always be applied
to transform the linear multiplet $L$ into a chiral
multiplet, leading
to a new equivalent chiral theory. Our goal in this
section is to extend
this equivalence (which will be called chiral-linear
duality) to the
effective theories describing in both cases gaugino condensates.

We have seen that a generic supersymmetric theory
with a linear multiplet,
defined by the lagrangian ${\cal L}_L = 2\mu^2
\int d^2\theta d^2\ov\theta\, \Phi(\hat L/\mu^2)$,
as in eq. (\ref{LL1}), can be first transformed
into a theory with a vector
superfield $V$ replacing $\hat L$ and a chiral superfield
${\cal S}$ used as a Lagrange multiplier to enforce
this replacement. This theory,
which is given in eq. (\ref{LL2}) will be
the starting point of the duality analysis.
Lagrangian (\ref{LL2}) can be rewritten
\beq
\label{dual1}
{\cal L}_V = \Dint\left[ 2\mu^2 \Phi(V/\mu^2) - ({\cal S}+\ov{\cal S})V
\right]
+\left({1\over2}\sum_ac_a\Fint {\cal S}\Tr(W_aW_a) +{\rm h.c.}\right).
\eeq
(This step involves a space-time integration by parts and a space-time
derivative is dropped.) Integrating over ${\cal S}$
leads back to expression
(\ref{LL1}), the theory with the linear superfield. Alternatively,
integrating over the vector superfield $V$ leads to a theory with chiral
${\cal S}$ and gauge superfields only (as well as the
omitted gauge-singlet
matter superfields). The equation of motion for $V$,
\beq
\label{Veom}
2\Phi_x = {\cal S}+\ov{\cal S},
\eeq
[$\Phi_x$ is defined in eq. (\ref{g2})]
can be used (in principle) to express $V/\mu^2$ as a function of
${\cal S}+\ov{\cal S}$. This expression inserted into eq. (\ref{dual1})
leads to the dual theory
\beq
\label{dual2}
{\cal L}_{\cal S} = \mu^2 \Dint K({\cal S}+\ov{\cal S})
+\left({1\over2}\sum_ac_a\Fint {\cal S}\Tr(W_aW_a) +{\rm h.c.}\right),
\eeq
with a K\"ahler function $K$ which only depends on ${\cal S}+\ov{\cal S}$
and without ${\cal S}$-dependent superpotential.
Theory (\ref{dual2}) is of the form (\ref{LS1})
with a particular ${\cal L}_S$. Explicitly,
\beq
\label{Kis}
K({\cal S}+\ov{\cal S}) =\left[
2\Phi(x) - ({\cal S}+\ov{\cal S})x\right]
_{x=V({\cal S}+\ov{\cal S})/\mu^2}.
\eeq
This equation implies various identities. One of them is
\beq
\label{ident}
K_{{\cal S}\ov{\cal S}} =
{\partial^2\over\partial {\cal S}\partial\ov{\cal S}}
K({\cal S}+\ov{\cal S})
=-{1\over2}[\Phi_{xx}]^{-1},
\qquad\qquad
\Phi_{xx} = \left[{d^2\over dx^2}\phi(x)\right]_{
x=V({\cal S}+\ov{\cal S})/\mu^2},
\eeq
which indicates that the K\"ahler metric $K_{{\cal S}\ov{\cal S}}$
of the dual theory (\ref{dual2}) is proportional to the inverse
of the metric $\Phi_{xx}$ of the theory with the linear multiplet.

The field-dependent gauge couplings in the linear multiplet theory are
given by eq. (\ref{g2}). On the other hand, theory (\ref{dual2}) has
$$
{1\over g_a^2} = c_a ({\cal S}+\ov{\cal S})|_{\theta=0}.
$$
The equality of gauge couplings in both formulations is the lowest
component of the superfield equation (\ref{Veom}).

We are now ready to extend chiral-linear duality
to the effective lagrangian
for gaugino condensates. Using the results of section 2 for theory
(\ref{dual2}), one obtains
\beq
\label{LeffS}
\begin{array}{rcl}
{\cal L}_{eff.}({\cal S},U_a) &=& \displaystyle{\Dint\left[\mu^2 K({\cal
S}+\ov{\cal S})
+\sum_a h_a (U_a\ov U_a)^{1/3} \right] }\crbig
&& \displaystyle{+ \left( \sum_a\Fint \left[{1\over2}c_a{\cal S}U_a
+{1\over12}A_aU_a\log(U_a/M^3) \right]
+{\rm h.c.} \right)}.
\end{array}
\eeq
And eq. (\ref{LeffL1}) can be rewritten
\beq
\label{LeffL}
\begin{array}{rcl}
{\cal L}_{eff.}(V,U_a) &=& \displaystyle{
\Dint\left[2\mu^2\Phi(V/\mu^2) -({\cal S}+\ov{\cal S})V  +\sum_a h_a
(U_a\ov U_a)^{1/3} \right] }\crbig
&& \displaystyle{+ \left( \sum_a\Fint \left[{1\over2}c_a{\cal S}U_a
+{1\over12}A_aU_a\log(U_a/M^3) \right]
+{\rm h.c.} \right)}.
\end{array}
\eeq
The duality equations (\ref{Veom}) and (\ref{Kis}),
which allow to exchange
$V/\mu^2$ and $\Phi$ with ${\cal S}+\ov{\cal S}$ and
$K$ manifestly display
the equivalence of these two expressions for ${\cal L}_{eff.}$.

It should be clear that in the second form of the effective lagrangian,
${\cal S}$ is a Lagrange multiplier which imposes the constraint
(\ref{const}). ${\cal L}_{eff.}$ has then
variables $V$ and $U_a$, submitted
to this condition. At the level of effective lagrangians,
chiral-linear duality
becomes a duality exchanging
$$
{\cal S} \qquad\longleftrightarrow\qquad V \quad{\rm with}\quad
\ov{\cal DD}V = -2\sum_a c_aU_a.
$$
Suppose we decompose the real vector superfield $V$ according to
$$
V = V_L +\Delta V, \qquad\qquad {\cal DD}V_L =\ov{\cal DD}V_L = 0.
$$
$V_L$ is a linear multiplet with four bosonic and four fermionic (4B+4F)
off-shell components, and closed supersymmetry
transformations. $\Delta V_L$
also contains (4B+4F) off-shell components,
as well as ${\cal S}$ and each
$U_a$. Clearly, the equation $\ov{\cal DD}V = -2\sum_a c_aU_a$
does not act on the linear multiplet component $V_L$ of $V$.
It eliminates
(4B+4F) components in the set $\{\Delta V,U_a\}$. Considering component
fields, it is tempting to conclude that
chiral-linear, $L\longleftrightarrow{\cal S}$ duality of the microscopic
theory simply leads to $V_L\longleftrightarrow{\cal S}$ duality
in effective lagrangians. But the important point is that supersymmetry
imposes to introduce the entire vector superfield
$V$ in ${\cal L}_{eff.}$
simply because condition $\ov{\cal DD}\Delta V = -2\sum_a c_aU_a$
cannot be used to eliminate $\Delta V$ as a (local)
function of the $U_a$'s.

We have then proven in the effective lagrangian approach
that chiral-linear duality does survive gaugino condensation, but
the linear multiplet $L$ is embedded in a
full real vector superfield $V$
with constraint (\ref{const}).

\section{One gaugino condensate and symmetries}

We begin the analysis of gaugino condensates with the
case of a simple
gauge group. The two dual effective
lagrangians are given by eqs. (\ref{LeffS})
and (\ref{Leffsimple}).
We will here only be interested in bosonic contributions.
A more complete discussion can be found in ref. \cite{BDQQ1}.

In the more familiar chiral multiplet formulation,
the effective lagrangian depends on two chiral superfields
${\cal S}$ et $U$. Each contains a
complex scalar ($s$ and $u$) and a complex auxiliary field
($f_s$ and $f_u$). The effective lagrangian
describes then four propagating bosonic degrees of
freedom. The real scalar field $\Re s$ defines the gauge coupling
[see eq.
(\ref{gaugeS})] and replaces the dilaton.
Expectation values of gaugino condensates are described by
$u$. The auxiliary field lagrangian is
\beq
\label{LauxS}
\begin{array}{rcl}
{\cal L}_{aux.} &=& {h\over9}(u\ov u)^{-2/3} f_u\ov f_u
+\mu^2 K_{s\ov s}f_s \ov f_s
-{1\over2}f_s u - {1\over2}\ov f_s\ov u \crbig
&& -\Re f_u\left(\Re s +{A\over12} \left[ 2+\log(u\ov u/M^6)\right]
\right)
+\Im f_u\left(\Im s-i{A\over12} \log(u/\ov u) \right),
\end{array}
\eeq
and the scalar potential reads
\beq
\label{VS}
\begin{array}{rcl}
V &=&\displaystyle{
{9\over4h}(u\ov u)^{2/3} \left\{ \Re s +{A\over12} [2+
\log({u\ov u\over M^6}) ] \right\}^2 + {1\over4}\mu^{-2}
K_{s\ov s}^{-1}u\ov u } \crbig
&&\displaystyle{
+{9\over4h}(u\ov u)^{2/3}\left\{\Im s +i{A\over12}
\log\left({\ov u\over u}\right) \right\}^2. }
\end{array}
\eeq
Actually, apart from kinetic terms $\mu^2K_{s\ov s}(\partial_\mu\Im s)
(\partial^\mu\Im s)$, the bosonic lagrangian only depends on $\Im s$
through the last term in (\ref{VS}). The effective lagrangian is
exactly invariant
under the combination of transformation (\ref{Ssym}) and the R-symmetry
which rotates the phase of the gaugino condensate,
$ u \longrightarrow e^{-6i\alpha/A}u$. More precisely,
the effective lagrangian
(\ref{LeffS}) is anomalous under each of these two transformations: its
variation is a space-time derivative. For (\ref{Ssym}),
this follows from eq. (\ref{Va}):
$$
\delta \big(\Fint {\cal S}U +{\rm h.c.}\big) =
\alpha \Im\big( \Fint \ov{\cal DD}V \big) = \partial^\mu(\ldots).
$$
And the combined transformation is anomaly-free. This
symmetry indicates that for each choice of condensate phase,
there exists
a value for $\langle\Im s\rangle$ which
minimizes $V$, and physics does not
depend on this value. Gaugino condensation
does not break the shift symmetry
(\ref{Ssym}) which is essential for chiral-linear duality.

The first term in the potential (\ref{VS})
specifies the size of the gaugino condensate
as a function of the gauge coupling,
\beq
\label{minequ}
|u| = M^3e^{-1}e^{-6\langle\Re s\rangle/A} =
M^3 e^{-1} {\rm exp}\,\left(-{8\pi^2\over C(G)g^2}\right),
\eeq
while the second one leads to the standard runaway behaviour towards the
weak coupling limit $u\longrightarrow 0$ which is a well-known feature of
global supersymmetry. These last results would be affected by the coupling
to gravity, but the symmetry which combines R-rotation of the condensate
and shift transformation (\ref{Ssym}) will survive supergravity effects.

In the linear multiplet formulation, the effective lagrangian is given
by eq. (\ref{Leffsimple}). It depends on a vector
superfield $V$ and on a single chiral $U$, related by $U=-{1\over2}\ov
{\cal DD}V$, according to condition (\ref{const}).
The bosonic parts of $V$ and $U$ can be written
\beq
\label{UVcomp}
\begin{array}{rcl}
V &=& c-{1\over2}\theta\theta\ov m -{1\over2}\ov{\theta\theta}m
-\theta\sigma^\mu\ov\theta v_\mu
+\theta\theta\ov{\theta\theta}(d+{1\over4}\Box c), \crbig
U &=& u -i\theta\sigma^\mu\ov\theta \partial_\mu u
-{1\over4}\theta\theta
\ov{\theta\theta}\Box u -\theta\theta f_u,
\end{array}
\eeq
with
\beq
\label{UVrel}
u = -m,
\qquad\qquad
f_u = -2d+ i\partial^\mu v_\mu.
\eeq
Gaugino condensates correspond to expectation values
of $u=-m$, the real scalar
$c$ is the dilaton and $v_\mu$ should be identified with
$$
{1\over\sqrt2}\epsilon_{\mu\nu\rho\sigma}h^{\nu\rho\sigma}
={1\over\sqrt2}\epsilon_{\mu\nu\rho\sigma}
\left(\partial^\nu b^{\rho\sigma}
+\sqrt2 \omega^{\nu\rho\sigma}\right) -
\Tr(\lambda\sigma^\mu\ov\lambda).
$$
Actually $V$ can as well be regarded as the supersymmetrization of the
antisymmetric tensor $h^{\nu\rho\sigma}$, and
the chiral superfield $U$ as
the supersymmetric extension of its curl \cite{G}. After eliminating
the real auxiliary $d$, the bosonic effective lagrangian is
\beq
\label{Lbos}
\begin{array}{rcl}
{\cal L}_{bos.} &=& \displaystyle{
- {1\over2}\mu^{-2}\Phi_{xx}\left[
(\partial_\mu c)(\partial^\mu c) - v^\mu v_\mu \right]
+{h\over9} (u\ov u)^{-2/3} \left[ (\partial_\mu \ov u)(\partial^\mu u)
+ (\partial_\mu v^\mu)^2 \right]  }\crbig
&& \displaystyle{
+ {A\over12}i(\partial^\mu v_\mu) \log \left({\ov u\over u}\right) - V.}
\end{array}
\eeq
[The notation is as in eq. (\ref{ident})]. The four
propagating degrees of freedom are $c$ (dilaton),
$u$ (condensate, complex field) and one component of the vector
$v_\mu$ (the longitudinal part). The four auxiliary bosons are $d$ and
three components of $v_\mu$. In comparison with the chiral case,
$c$ and the longitudinal part of $v_\mu$ replace $s$ [see
eq. (\ref{Veom})]. In eq. (\ref{Lbos}), the
quantities $-{1\over2}\mu^{-2}\Phi_{xx}$ and
${h\over9} (u\ov u)^{-2/3}$ should be positive to ensure boundedness of
kinetic terms. The scalar potential is
\beq
\label{pot}
V =
{9\over16h}(u\ov u)^{2/3}\left\{ 2\Phi_x+{A\over6}\left[2+
\log\left({u\ov u\over M^6}\right)\right]\right\}^2
-{1\over2}\mu^{-2}\Phi_{xx} u\ov u.
\eeq
As it should with global supersymmetry, it is the sum
of two positive terms.
As before, rotating the condensate phase by an angle $\beta$
leaves physics unchanged since $\delta[{\cal L}_{bos.}] = \beta{A\over6}
(\partial^\mu v_\mu)$: the effective lagrangian does not specify the
phase of the condensate. And minimizing the potential leads again to eq.
(\ref{minequ}) and to the runaway behaviour.

\section{Several condensates and conclusions}

The analysis of the effective lagrangian for a semi-simple gauge group
and in the chiral formulation, eq. (\ref{LeffS}), is simple and familiar
\cite{krasnikov}. Minimizing the potential
leads to the following equations:
\beq
\label{sev1}
\begin{array}{rcl}
u_a &=& \displaystyle{
M^3e^{-1}e^{-6c_a s/A_a} = M^3e^{-1}e^{-16\pi^2 s c_a/C(G_a)},
\qquad\qquad a=1,2,\ldots,} \crbig
\displaystyle{\sum_a c_a u_a} &=& 0.
\end{array}
\eeq
Again $\langle\Im s\rangle$ is directly related to the phases of
gaugino condensates $u_a$, but the second condition
is not invariant under
a shift of $\langle\Im s\rangle$ except if all condensates vanish.
The shift
symmetry (\ref{Ssym}) is then broken with several condensates.

The analysis of the effective lagrangian in the linear formulation, eq.
(\ref{LeffL}), requires more care. One has to take the superfield
constraint $-2\sum_c c_aU_a = \ov{\cal DD}V$ into account. At the bosonic
level, it translates into
\beq
\label{sev2}
\sum_a c_au_a = -m \qquad{\rm and}\qquad \sum_c c_a f_{u_a}
=-2d+i\partial^\mu v_\mu,
\eeq
which cannot be used to eliminate the auxiliary fields $f_{u_a}$. A
Lagrange multiplier field is necessary. As
explained in ref. \cite{BDQQ1}, the minimization of
the potential with these constraints leads firstly
to the second condition
(\ref{sev1}). This is simply the statement that
unbroken supersymmetry requires
$m=0$ because $m$ is the $\ov{\theta\theta}$ component of $V$.
It secondly leads to
$$
u_a = M^3e^{-1} e^{-6c_a(\Phi_x-2i\xi)/A_a},
$$
with an arbitrary phase $\xi$. Since $\Phi_x=\Re s$ by duality, this is
again the first eq. (\ref{sev1}), and both formulations lead,
as expected, to the same minimum structure.

In the chiral case, the breaking of the shift symmetry (\ref{Ssym})
generates a mass for the axion $\Im s$. The dual version
has a vector field
$v_\mu$ and no axion. The computation of
the kinetic lagrangian for $v_\mu$ in the vacuum defined
by the minimum of the potential shows
that only one component of $v_\mu$ propagates, as in the
single gaugino case. One also verifies that this propagating
degree of freedom gets a
mass which is identical to the axion mass in the chiral case.
The linear formulation provides then a lagrangian description of a massive
axion in terms of a vector field or, equivalently, in terms of a
massive three-index antisymmetric tensor field.

These results show that there exists a formulation of
gaugino condensation
entirely constructed in the linear multiplet formalism.
This permits the
analysis to be performed using the multiplet
in which string theory presents
the dilaton. Chiral-linear duality, which is
well established at the (string)
perturbative level, admits an extension in the
presence of nonperturbative
gaugino condensates. The extension involves
non-trivial couplings of a
real vector superfield, replacing the linear
multiplet in the effective lagrangian. The axion of the
chiral formulation is replaced by an antisymmetric tensor which acquires
a mass in exactly the same way as the axion.

\end{document}